\documentclass[%
 reprint,
superscriptaddress,
 amsmath,amssymb,
 aps,
]{revtex4-2}

\usepackage{graphicx}% Include figure files
\usepackage{dcolumn}% Align table columns on decimal point
\usepackage{bbold}
\usepackage[version=3]{mhchem}
\usepackage{siunitx}
\usepackage{hyperref}
\begin{document}

\title{All-electron many-body approach to resonant inelastic x-ray scattering}% Force line breaks with \\

\author{Christian Vorwerk }
  \thanks{current address: Pritzker School of Molecular Engineering, University of Chicago, Chicago, IL 60637, USA.}
 \affiliation{Physics Department and IRIS Adlershof, Humboldt-Universit\"at zu Berlin, Zum Gro\ss en Windkanal 2, Berlin, Germany. }

\author{Francesco Sottile}%
  \affiliation{LSI, Ecole Polytechnique,     
  CNRS, CEA, Institut Polytechnique de Paris, F-91128 Palaiseau, France.}%

\author{Claudia Draxl}
 \email{claudia.draxl@physik.hu-berlin.de}
 \affiliation{Physics Department and IRIS Adlershof, Humboldt-Universit\"at zu Berlin, Zum Gro\ss en Windkanal 2, Berlin, Germany. }

\date{\today}% It is always \today, today,
             %  but any date may be explicitly specified

\begin{abstract}
We present a formalism for the resonant inelastic x-ray scattering (RIXS) cross
section. The resulting compact expression in terms of polarizability matrix
elements, particularly lends itself to the implementation in an all-electron
many-body perturbation theory (MBPT) framework, which is realized in the
full-potential package \texttt{exciting}. With the carbon K edge RIXS of diamond and the
oxygen K edge RIXS of $\beta-\ce{Ga2O3}$, respectively, we demonstrate the
importance of electron-hole correlation and atomic coherence in the RIXS
spectra.
\end{abstract}

\maketitle

%------------------------------------------------------------------------------
\section{Introduction}
%------------------------------------------------------------------------------
Resonant inelastic x-ray scattering (RIXS) is a "photon-in-photon-out"
scattering process, consisting of a coherent x-ray absorption and x-ray emission
process.\cite{Kotani2001,Ament2011} The energy difference between the absorbed
and emitted x-ray photons is transferred to the system. Being bulk-sensitive,
element- and orbital-specific and giving access to a large scattering phase
space, RIXS has become a widely used experimental probe of elementary
excitations in molecules~\cite{cesar1997,hennies2007,josefsson2012} and
solids.\cite{Kotani2001,Ament2011,ghiringhelli2005,wang2017} Accurate \textit{ab
initio} simulations can provide insight into the complex RIXS process and
predict the corresponding spectra. In this context, the electron-hole
correlations of the x-ray absorption and emission processes as well as the
quantum coherence between them is a challenge for {\it ab initio} approaches.

For the first-principles calculations of both valence and core excitations in
solids, the Bethe-Salpeter equation (BSE)
formalism~\cite{hedin1965,hybe-loui85prl,Strinati1988,onida1995,benedict1998,rohlfing1998}
has become the state of the art in recent decades. However, only few
applications of the BSE formalism to
RIXS~\cite{shirley2000,shirley2000a,shirley2001,johnvinson2012,Vinson2017,vinson2019,Geondzhian2018}
have been presented so far. All of them relied on the pseudopotential
approximation where only selected valence and conduction orbitals are explicitly
included in the calculation, while more strongly-bound electrons are only
treated implicitly through the pseudopotential. The wavefunctions required for
the calculations of x-ray absorption and scattering in these BSE implementations
\cite{Shirley1998,shirley2000,gilmore2015,shirley2000,Vinson2016,Vinson2017}rely
on the pseudopotential approximations via the projector augmented-wave
method.\cite{carlisle1999a,vinson2011,blochl1994,holzwarth2001}

In this work, we go beyond by presenting an all-electron full-potential BSE
approach to RIXS. We derive a compact analytical expression of the RIXS cross
section within many-body perturbation theory (MBPT) that contains both the
effects of electron-hole correlation and of the quantum coherence of the
resonant scattering. We have implemented this novel expression in the
all-electron many-body perturbation theory code
\texttt{exciting},\cite{gulans2014,vorwerk2017,vorwerk2019} where we make use of
consistent BSE calculations of core and valence excitations, based on the
explicit access to core, valence, and conduction orbitals and matrix elements
between them. To demonstrate our approach, we have chosen two representative
examples: In the carbon K edge of diamond, we particularly study the influence
of electron-hole correlation, while in the oxygen K edge of $\beta-\ce{Ga2O3}$,
we show how the coherence between excitations at different atomic sites impacts
the RIXS spectra.

%------------------------------------------------------------------------------
\section{Theory}
%------------------------------------------------------------------------------
The double-differential cross section (DDCS) $\mathrm{d}^2 \sigma/\mathbf{d}
\Omega_2 \mathrm{d} \omega_2$ for the scattering of a photon with energy
$\omega_1$, polarization $\mathbf{e}_1$, and momentum $\mathbf{K}_1$ into a
photon with energy $\omega_2$, polarization $\mathbf{e}_2$, and momentum
$\mathbf{K}_2$ is given by the generalized Kramers-Heisenberg \cite{Dirac1927a}
formula
\begin{equation}\label{eq:hkf}
    \begin{aligned}
  \frac{\mathrm{d}^2 \sigma}{\mathrm{d} \Omega_2 \mathrm{d}\omega_2}
  &=\alpha^4 \left(\frac{\omega_2}{\omega_1} \right)\sum_f \left| \langle f |
  \mathbf{e}_1 \cdot \mathbf{e}_2^*
  \sum_j \mathrm{e}^{\mathrm{i}\mathbf{Q}\mathbf{r}_j}| 0 \rangle \right.\\ 
  &\left.+\sum_n \frac{\langle f | \sum_j 
  \mathrm{e}^{-\mathrm{i}\mathbf{K}_2\mathbf{r}_j}\mathbf{e}_2^* 
  \cdot\mathbf{p}_j | n \rangle \langle n | 
  \mathrm{e}^{\mathrm{i}\mathbf{K}_1\mathbf{r}_j}
  \mathbf{e}_1 \cdot \mathbf{p}_j | 0\rangle}
  {\omega_1-E_n}\right|^2 \\
  & \times \delta((\omega_1-\omega_2)-E_f),
\end{aligned} 
\end{equation}
where $| 0 \rangle$ and $| f \rangle$ are the initial and final electronic state
with energy $E_0=0$ and $E_f$, respectively. Equation~\ref{eq:hkf} includes both
the non-resonant scattering (NRIXS), in which the momentum transfer
$\mathbf{Q}=\mathbf{K}_1-\mathbf{K}_2$ appears explicitly, and the resonant
scattering (RIXS). If the photon energy $\omega_1$ is in resonance with the
excitation energies $E_n$ of the electronic system, the second, resonant term
dominates and the non-resonant part can be neglected, leading to
\begin{equation}\label{eq:hks2}
\begin{aligned}
      \frac{\mathrm{d}^2 \sigma}{\mathrm{d} \Omega_2 \mathrm{d}\omega_2}
  =&\alpha^4 \left( \frac{\omega_2}{\omega_1} \right) \sum_f
  \bigg|\sum_n \frac{\langle f |\hat{T}^{\dagger}(\mathbf{e}_2) 
  | n \rangle \langle n | 
  \hat{T}(\mathbf{e}_1) | i \rangle}
  {\omega_1-E_n+\mathrm{i}\eta}\bigg|^2\\ &\times \delta(\omega_1-\omega_2 - E_f),
  \end{aligned}
\end{equation}
where we have introduced the transition operator $\hat{T}(\mathbf{e})$ in the
dipole approximation, \textit{i.e.} $\hat{T}(\mathbf{e}) \approx \sum_j \mathbf{e} \cdot
\mathbf{p}_j$. Microscopically, the resonant scattering process is as follows:
The absorption of the initial x-ray photon with energy $\omega_1$ excites a core
electron into the conduction band, leaving a core hole behind. The intermediate
system, after the excitation, is in an excited many-body state $| n \rangle$.
The core hole can now be filled by a valence electron, which loses the energy
$\omega_2$ by emitting an x-ray photon. This process is known as \textit{direct}
RIXS. Alternatively, the presence of the core-hole and excited electron can lead
to the formation of a secondary electron-hole pair in the intermediate state.
The initially excited can then fill the core hole. This process is known as
\textit{indirect} RIXS.\cite{Ament2011,vanveenendaal2011} Involving excited
states beyond singlet excitations, its contribution is neglected in the
following. We recall that indirect RIXS is mostly devoted to magnetic
excitations, like magnon scattering.

In the final many-body state $|f \rangle$, a hole is present in a valence state
and an excited electron in a conduction state. While both the absorbed and
emitted photon have energies in the x-ray region, the difference between them,
the energy loss $\omega_1-\omega_2$, is typically in the range of several eV.
The final excited many-body state of the RIXS process corresponds to an excited
state as typically created by optical absorption. The intricacy of the RIXS
formalism, described in Eq.~\ref{eq:hks2}, arises from the transitions between
three many-body states, \textit{i.e.} $| 0 \rangle \rightarrow | n \rangle$ and $| n
\rangle \rightarrow | f \rangle$ and from the coherent summation over all
possible intermediate and final many-body states. This inherent complexity of
the microscopic RIXS process poses challenges for any theoretical description,
as both the effects of electron-hole interaction, as well as the coherence of
the RIXS process have to be included.

%------------------------------------------------------------------------------
\subsection{Independent-Particle Approximation to RIXS}
%------------------------------------------------------------------------------
It is instructive to discuss the RIXS process within the independent-particle
approximation (IPA) before the scattering in the fully interactive system is
considered. In the IPA, the many-body groundstate wavefunction is given by a
single Slater determinant, and both the intermediate many-body state $|n\rangle$
and the final one $|f\rangle$ are singlet excitations of the groundstate without
any relaxation of the system. We know \textit{a priori} that the intermediate
states contain a core hole $\mu \mathbf{k}$ and an excited electron in a
conduction state $c \mathbf{k}$, such that we can express them in second
quantization as 
$| n \rangle = |c \mu \mathbf{k} \rangle=\hat{c}^{\dagger}_{c
\mathbf{k}}\hat{c}_{\mu \mathbf{k}}| 0 \rangle$ with energy $E_n=\epsilon_{c
\mathbf{k}}-\epsilon_{\mu \mathbf{k}}$.  Furthermore, the final states contain
an excited electron in a specific conduction state $c' \mathbf{k}'$ and a
valence hole in the state $v \mathbf{k}'$, such that $| f \rangle =
|c'\mathbf{k}' v\mathbf{k}' \rangle= \hat{c}^{\dagger}_{c'\mathbf{k}'}\hat{c}_{v
\mathbf{k}'}| 0 \rangle$ and $E_f=\epsilon_{c' \mathbf{k}'}-\epsilon_{v
\mathbf{k}'}$. Then, Eq.~\ref{eq:hks2} becomes

\begin{equation}\label{eq:ip-start}
\begin{aligned}
  &\frac{\textrm{d}^2 \sigma}{\textrm{d}\Omega_2\textrm{d}\omega_2} \bigg|_{\mathrm{IP}} 
  =\\
  &\alpha^4 \left( \frac{\omega_2}{\omega_1} \right)
  \underbrace{\sum_{c' v \mathbf{k}'}}_{f}  \Big| \underbrace{\sum_{c \mu \mathbf{k}}}_{n} 
  \frac{\langle c' v \mathbf{k}' | \hat{T}^{\dagger}(\mathbf{e}_2) |
  c \mu \mathbf{k} \rangle \langle   c \mu \mathbf{k} |   \hat{T}(\mathbf{e}_1) | 0 \rangle}
  {\omega_1-(\epsilon_{c \mathbf{k}}-\epsilon_{\mu \mathbf{k}})   +\textrm{i}\eta} \Big|^2   \times\\
  & \times \delta(\omega-(\epsilon_{c'\mathbf{k}'}   -\epsilon_{v \mathbf{k}'})),
\end{aligned}
\end{equation}
In second quantization, we express the transition operator as
$\hat{T}=\sum_{mn}\sum_{\mathbf{k}}\mathbf{e}_1 \cdot \mathbf{P}_{mn \mathbf{k}}
\hat{c}^{\dagger}_{m \mathbf{k}}\hat{c}_{n \mathbf{k}}$, where $\mathbf{P}_{mn
\mathbf{k}}=\langle m \mathbf{k}|\mathbf{p}| n \mathbf{k}\rangle$ are the
momentum matrix elements. Inserting these operators in Eq.~\ref{eq:ip-start}
yields
\begin{equation}\label{eq:rixs3}
  \begin{aligned}
    &\frac{\textrm{d}^2 \sigma}{\textrm{d}\Omega_2\textrm{d}\omega_2} 
    \bigg|_{\mathrm{IP}} =\\
    &\sum_{c' v \mathbf{k}'}  \bigg| \sum_{c \mu \mathbf{k}}
    \sum_{m n \mathbf{k}''} \sum_{p q \mathbf{k}'''}\left[
    \mathbf{e}^*_{2}\cdot \mathbf{P}_{m n \mathbf{k}''} \right] \times \\
    &\times\frac{\langle c' v  \mathbf{k}'| \hat{c}_{m \mathbf{k}''}^{\dagger}
    \hat{c}_{n \mathbf{k}''} |
    c \mu \mathbf{k} \rangle \langle   c \mu \mathbf{k} | 
    \hat{c}_{p \mathbf{k}'''}^{\dagger}\hat{c}_{q \mathbf{k}'''} | 0 \rangle}
    {\omega_1-(\epsilon_{c \mathbf{k}}-\epsilon_{\mu \mathbf{k}})
    +\textrm{i}\eta}  
    \left[\mathbf{P}_{p q \mathbf{k}'''}\cdot \mathbf{e}_1 \right] \Bigg| ^2
    \times \\ 
    & \times
    \delta(\omega-(\epsilon_{c' \mathbf{k}'}-\epsilon_{v \mathbf{k}'})).
  \end{aligned}
\end{equation}
We note that the summations over $p$ and $q$ in Eq.~\ref{eq:rixs3} are not
restricted to either core, valence, or conduction states. Restrictions to these
indices can be inferred from the matrix elements of the creation and
annihilation operators. We find that
\begin{equation}\label{eq:rixs4}
  \langle c \mu \mathbf{k} | 
  \hat{c}_{p \mathbf{k}'''}^{\dagger}\hat{c}_{q \mathbf{k}'''} | 0 \rangle
  =  \delta_{\mu q}\delta_{c p}
  \delta_{\mathbf{k} \mathbf{k}'''}.
\end{equation}
The term 
$\langle c' v \mathbf{k}' | \hat{c}_{m \mathbf{k}''}^{\dagger}\hat{c}_{n
\mathbf{k}''} |c \mu \mathbf{k} \rangle$ requires a more careful treatment.
Applying Wick's theorem and restricting only to terms that correspond to the
RIXS process, we obtain
\begin{equation}
\label{eq:rixs7}
  \langle c v \mathbf{k} |  \hat{c}_{m \mathbf{k}''}^{\dagger}
  \hat{c}_{n \mathbf{k}''} |c' \mu \mathbf{k}' \rangle =
  -\delta_{cc'}\delta_{\mu m} \delta_{v n}
  \delta_{\mathbf{k}\mathbf{k}'}\delta_{ \mathbf{k}\mathbf{k}''}.
\end{equation}
A more detailed derivation of Eq.~\ref{eq:rixs7} is provided in the Appendix.
Inserting Eqs.~\ref{eq:rixs4} and \ref{eq:rixs7} into Eq.~\ref{eq:rixs3} yields
\begin{equation}\label{eq4:rixs7-1}
\begin{aligned}
  \frac{\textrm{d}^2 \sigma}{\textrm{d}\Omega_2\textrm{d}\omega_2} 
  \bigg|_{\mathrm{IP}}
  =&\alpha^4 \left( \frac{\omega_2}{\omega_1}\right)
  \sum_{c v \mu \mathbf{k}}\left| 
  \frac{\mathbf{e}^*_2  \cdot \mathbf{P}_{\mu v \mathbf{k}}
  \mathbf{P}_{c \mu \mathbf{k}}\cdot \mathbf{e}_1}
  {\omega_1-(\epsilon_{c \mathbf{k}}-\epsilon_{\mu \mathbf{k}})+\mathrm{i}\eta}
  \right|^2 \times \\
  & \times \delta(\omega-(\epsilon_{c \mathbf{k}}-\epsilon_{v \mathbf{k}}))\\
  =&-\frac{\alpha^4}{\pi}
  \left(\frac{\omega_2}{\omega_1}\right)
  \mathrm{Im}\; \sum_{cv \mathbf{k}}\frac{\left| \sum_{\mu} 
  \frac{\mathbf{e}^*_2 \cdot \mathbf{P}_{\mu v \mathbf{k}}
  \mathbf{P}_{c \mu \mathbf{k}}\cdot \mathbf{e}_1}
  {\omega_1-(\epsilon_{c \mathbf{k}}-\epsilon_{\mu \mathbf{k}})+\mathrm{i}\eta}
  \right|^2} 
  {\omega-(\epsilon_{c \mathbf{k}}-\epsilon_{v \mathbf{k}})+\mathrm{i}\eta}.
\end{aligned}
\end{equation}
This equation has been widely applied to calculate the RIXS cross section in
solids.\cite{Ma1992,Johnson1994,Jia1996,Strocov2004,Strocov2005,nisikawa2010}

%------------------------------------------------------------------------------
\subsection{RIXS beyond the Independent-Particle Approximation}
%------------------------------------------------------------------------------
The neglected electron-hole interaction is the reason for the poor performance
of the IPA for optical and x-ray excitation spectra in crystalline
semiconductors and insulators. A more accurate approach is provided by many-body
perturbation theory based on solutions of the
BSE.\cite{hedi65pr,Strinati1988,onida2002} This approach is now the state of the
art to determine
optical~\cite{hedi65pr,hybe-loui85prl,onida1995,albrecht1997,bene+98prl,rohl-loui98prl}
and x-ray absorption
spectra~\cite{vinson2011,vinson2012,noguchi2015,gilmore2015,cocchi2016,fossard2017,lask-blah10prb,vorwerk2017,drax-cocc17condmat,olovsson2009,olovsson2009a,olovsson2011,olovsson2013,cocchi2015a,vorwerk2018}
in solids. In the following, we derive an analytical expression for RIXS that
takes the electron-hole interaction into account.

Following Refs.\cite{shirley2000,johnvinson2012}, we define an intermediate
many-body state as
\begin{equation}\label{eq:y}
  | Y(\omega_1) \rangle = \sum_n \frac{| n \rangle \langle n |}{\omega_1
  -E_n}\hat{T}(\mathbf{e}_1)|0 \rangle.
\end{equation}
Similar intermediate states have been defined as \textit{response vectors} in
the context of non-linear spectroscopy in molecular
systems.\cite{nanda2017,nanda2020} Inserting these intermediate states into
Eq.~\ref{eq:hks2}, the RIXS cross section becomes
\begin{equation}\label{eq:ddcs_inter}
\begin{aligned}
  &\frac{\textrm{d}^2 \sigma}{\textrm{d}\Omega_2\textrm{d}\omega_2} =\\
  & \alpha^4 \left( \frac{\omega_2}{\omega_1} \right)
  \sum_f \langle Y(\omega_1) | \hat{T}(\mathbf{e}_2)|f\rangle
  \langle f| \hat{T}^{\dagger}(\mathbf{e}_2) | Y(\omega_1) \rangle \times \\
  & \times \delta(\omega-E_f).
\end{aligned}  
\end{equation}
The intermediate states $|Y(\omega_1)\rangle$ contain the information about all
possible excitation processes and can be understood as the excited many-body
states produced by the absorption of a photon with energy $\omega_1$.  Inserting
a finite broadening $\eta$ for the excited many-body states, the cross section
becomes
\begin{equation}\label{eq:ddcs_inter2}
  \frac{\textrm{d}^2 \sigma}{\textrm{d}\Omega_2\textrm{d}\omega_2}
  = \alpha^4 \left( \frac{\omega_2}{\omega_1} \right)
  \textrm{Im}\sum_{f} 
  \frac{\langle Y(\omega_1) |\hat{T}(\mathbf{e}_2)|f\rangle
  \langle f| \hat{T}^{\dagger}(\mathbf{e}_2) | Y(\omega_1) \rangle}
  {\omega-E_f+\textrm{i}\eta}.   
\end{equation}
Within the Tamm-Dancoff approximation, we assume that both the intermediate and
final states are linear combinations of singlet excitations of the groundstate.
We can thus infer
\begin{equation}\label{eq:id_tda}
  \mathbb{1} \approx \sum_{i \mathbf{k}}\sum_{j \mathbf{k}'} 
  \hat{c}^{\dagger}_{j \mathbf{k}'}
  \hat{c}_{i \mathbf{k}}| 0\rangle \langle 0 | \hat{c}^{\dagger}_{i \mathbf{k}}
  \hat{c}_{j \mathbf{k}'}.
\end{equation}
While this approximation has been found to yield good results for the excited
states probed in both optical and x-ray absorption spectroscopy of
solids,\cite{vorwerk2017,vorwerk2019} it explicitly limits our approach to
\textit{direct} RIXS, as the intermediate states in \textit{indirect} RIXS
contain two electron-hole pairs and therefore can not be expressed as a linear
combination of singlet excitations of the groundstate.
Inserting Eq.~\ref{eq:id_tda} into Eq.~\ref{eq:y} yields
\begin{equation}\label{eq:yw_chi}
  \begin{aligned}
    &|Y(\omega_1)\rangle=\\
    &\sum_n \sum_{c \mu \mathbf{k}}\sum_{c' \mu' \mathbf{k}'}
    \hat{c}^{\dagger}_{c \mathbf{k}} \hat{c}_{\mu \mathbf{k}}| 0\rangle 
    \frac{\langle 0 |  \hat{c}^{\dagger}_{\mu \mathbf{k}} \hat{c}_{c \mathbf{k}}
    |n \rangle \langle n | \hat{c}^{\dagger}_{c' \mathbf{k}'}
    \hat{c}_{\mu' \mathbf{k}'}| 0 \rangle}{\omega_1-E_n} 
    \left[ \mathbf{e}_1 \cdot \mathbf{P}_{c'\mu' \mathbf{k}'}\right]\\
    &=\sum_{c \mu \mathbf{k}}\sum_{c' \mu' \mathbf{k}'}
    \hat{c}^{\dagger}_{c \mathbf{k}} \hat{c}_{\mu \mathbf{k}} | 0 \rangle
    \chi_{c \mu \mathbf{k},c'\mu' \mathbf{k}'}(\omega_1) \left[\mathbf{e}_1 \cdot 
    \mathbf{P}_{c'\mu' \mathbf{k}'} \right],
  \end{aligned}
\end{equation}
where we have made use of the Lehmann representation of the polarizability
$\chi(\omega)$. We can now evaluate the expectation value of the intermediate
state
\begin{equation}\label{eq4:yw_explicit}
  \begin{aligned}
    &\langle Y(\omega_1) | \hat{c}^{\dagger}_{j \mathbf{k}''}
    \hat{c}_{i \mathbf{k}''} \hat{c}^{\dagger}_{c''' \mathbf{k}'''}
    \hat{c}^{\dagger}_{v''' \mathbf{k}'''} | 0\rangle\\
    &=-\sum_{c \mu \mathbf{k}}\sum_{c' \mu' \mathbf{k}'}
    \chi^*_{c \mu \mathbf{k},c' \mu' \mathbf{k}'}(\omega_1)
    \left[\mathbf{e}_1 \cdot \mathbf{P}_{c' \mu' \mathbf{k}'}\right]^*
    \delta_{cc'''}\delta_{v'''j}\delta_{\mu i}
    \delta_{\mathbf{k}\mathbf{k}'''}\delta_{\mathbf{k}\mathbf{k}''}
  \end{aligned}
\end{equation}
where we have used Eq.~\ref{eq:rixs7}. This eventually yields the
double-differential cross section as
\begin{equation}\label{eq:final}
  \begin{aligned}
    &\frac{\textrm{d}^2 \sigma}{\textrm{d}\Omega_2\textrm{d}\omega_2}
    = \alpha^4 \left( \frac{\omega_2}{\omega_1} \right) \textrm{Im} 
    \sum_{c, c', c'', c'''} \sum_{\mu, \mu', \mu '', \mu '''}
    \sum_{v,v'}\sum_{\mathbf{k}\mathbf{k}'\mathbf{k}''\mathbf{k}'''}\\
    &\left[ \left[ \mathbf{e}^*_2 \cdot \mathbf{P}_{\mu v \mathbf{k}}\right]
    \chi_{c \mu \mathbf{k}, c' \mu' \mathbf{k}'}(\omega_1) 
    \left[ \mathbf{e}_1 \cdot \mathbf{P}_{c'\mu' \mathbf{k}'}\right]\right]^* \times \\
    & \times 
    \chi_{c v \mathbf{k},c'' v' \mathbf{k}''}(\omega)
    \left[ \left[ \mathbf{e}_2^* \cdot 
    \mathbf{P}_{\mu'' v' \mathbf{k}''}\right] 
    \chi_{c'' \mu'' \mathbf{k}'', c''' \mu'''\mathbf{k}'''}(\omega_1) \right.
    \times \\
    & \left. \times
    \left[\mathbf{e}_1 \cdot \mathbf{P}_{c''' \mu''' \mathbf{k}'''}\right]
    \right].
  \end{aligned}
\end{equation}

Equation~\ref{eq:final} represents a main result of this paper: The RIXS cross
section is expressed solely in terms of the polarizability $\chi$. The
polarizability is evaluated twice, once at the x-ray excitation energy
$\omega_1$, and once at the energy loss $\omega=\omega_1-\omega_2$. The
cumbersome summations over all intermediate and final many-body states are thus
included in the polarizability, and Eq.~\ref{eq:final} avoids any explicit
summations over many-body states. 
%************************************************
\begin{figure*}[t]
  \centering
  \includegraphics[width=0.75\textwidth]{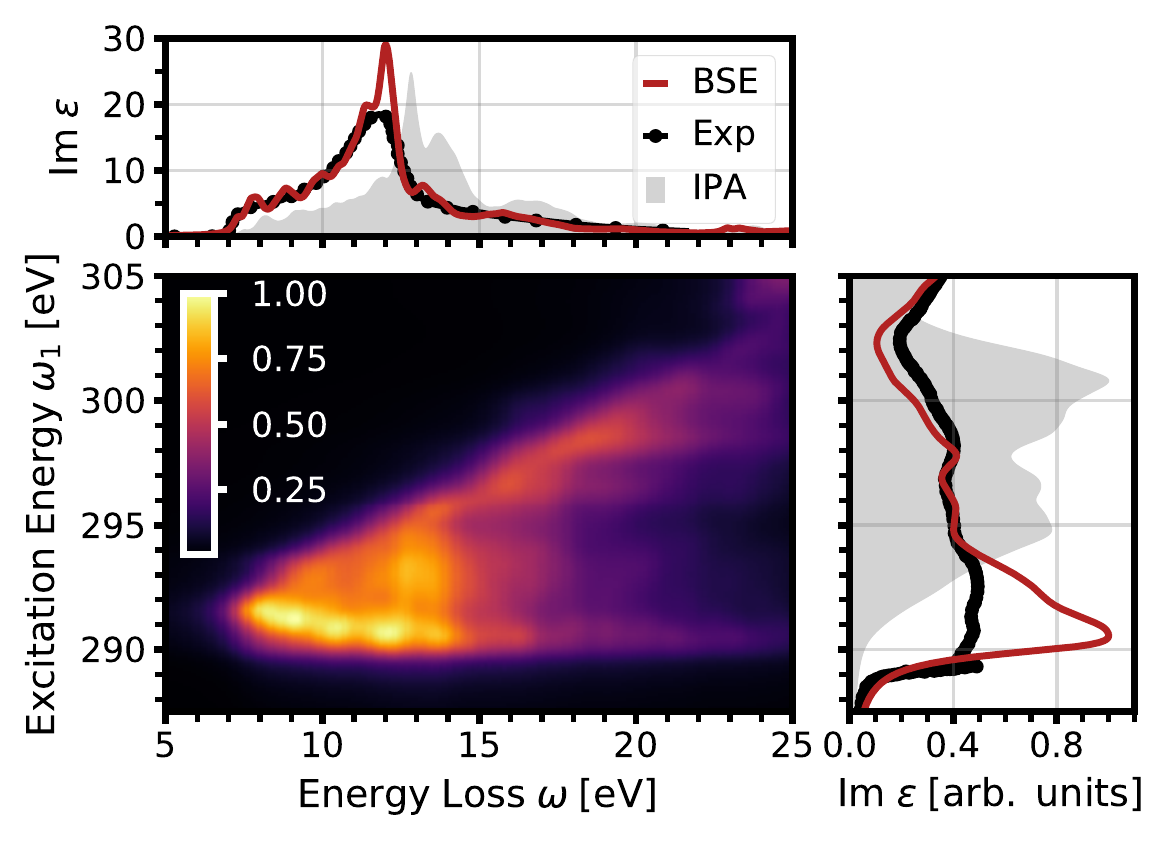}
  \caption{\label{fig:rixs-diamond1} Normalized double-differential RIXS cross
section of the carbon K edge in diamond as a function of excitation energy and
energy loss (center). On the top, the optical absorption spectrum obtained from
BSE (red) and the IPA (gray) is shown, compared to the experimental spectrum
(black) from Ref.~\cite{phillip1964} On the right, the corresponding core
excitation is depicted.}
\end{figure*}
%************************************************
%------------------------------------------------------------------------------
\subsection{Many-Body Perturbation Theory Applied to RIXS}
%------------------------------------------------------------------------------
Within MBPT, the polarizability is given by
\begin{equation}
\label{eq:chi_def}
  \chi_{ij \mathbf{k}, i'j' \mathbf{k}'}(\omega)=
  \sum_{\lambda} \frac{\left[X_{ij \mathbf{k},\lambda}\right]^*
  X_{i'j' \mathbf{k}',\lambda}}{\omega -E^{\lambda} + \mathrm{i}\eta},
\end{equation}
where $X_{ij \mathbf{k}, \lambda}$ and $E^\lambda$ are the eigenstates and
-values of the BSE  equation
\begin{equation}
  H^{\mathrm{BSE}} X_{\lambda} = E^\lambda X_\lambda,
\end{equation}
where the Hamiltonian is given by
\begin{equation}
  H^{\mathrm{BSE}} = \Delta E^{IP} + 2 V - W.
\end{equation}
Here, $\Delta E^{IP}$ are the single-particle energy differences which are taken
from DFT Kohn-Sham calculations and are corrected by a scissors operator. The
corresponding values for conduction states and core states are chosen such to
simulate the zero-order polarizability in the BSE calculations.\cite{Martin2016}
$V$ is the matrix elements of the bare electron-hole exchange, and $W$ those of
the statically screened direct interaction.

Inserting Eq.~\ref{eq:chi_def} into Eq.~\ref{eq:final} allows for a significant
simplification of the RIXS cross section. We first consider the last bracket of
Eq.~\ref{eq:final} and express it as
\begin{equation}\label{eq7:reform1}
  \begin{aligned}
    &\sum_{c''' \mu''' \mathbf{k}'''}\sum_{\mu''}
    \left[ \mathbf{e}_2^* \cdot \mathbf{P}_{\mu'' v' \mathbf{k}''}\right] 
    \chi_{c'' \mu'' \mathbf{k}'', c''' \mu'''\mathbf{k}'''}(\omega_1) \times\\
    &\times
    \left[\mathbf{e}_1 \cdot \mathbf{P}_{c''' \mu''' \mathbf{k}'''}\right]\\
    = &\sum_{\mu''}\sum_{\lambda_c}
    \left[ \mathbf{e}_2^* \cdot \mathbf{P}_{\mu'' v' \mathbf{k}''} \right]
    \frac{\left[ X_{c'' \mu'' \mathbf{k}'', \lambda_c} \right]^*
    t^{(1)}_{\lambda_c}}{\omega_1 - E^{\lambda_c}+\mathrm{i}\eta},
  \end{aligned}
\end{equation}
where $X_{c'' \mu'' \mathbf{k}''}$ and $E^{\lambda_c}$ are the eigenvectors and
-values of the core-level BSE, respectively. We define the
\textit{core-excitation oscillator strength} $t^{(1)}_{\lambda_c}$ as 
\begin{equation}\label{eq7:t1_def}
  t^{(1)}_{\lambda_c}=\sum_{c''' \mu''' \mathbf{k}'''}
  X_{c''' \mu''' \mathbf{k}''',\lambda_c}
  \left[\mathbf{e}_1 \cdot \mathbf{P}_{c''' \mu''' \mathbf{k}'''}\right]
\end{equation}
and the \textit{excitation pathway} $t^{(2)}_{\lambda_o,\lambda_c}$ as 
\begin{equation}
\label{eq7:t2_def}
  t^{(2)}_{\lambda_o,\lambda_c}=\sum_{cv \mathbf{k}}\sum_{\mu} 
  X_{cv \mathbf{k},\lambda_o}
  \left[ \mathbf{e}_2^* \cdot \mathbf{P}_{\mu v \mathbf{k}} \right]
  \left[ X_{c \mu \mathbf{k}, \lambda_c}\right]^*,
\end{equation}
where $X_{cv\mathbf{k},\lambda_{o}}$ and $E^{\lambda_o}$ are the eigenvectors
and -values of the BSE Hamiltonian of the valence-conduction transitions,
respectively. Here, we discern the index of the valence-conduction excitations,
$\lambda_o$, from the index $\lambda_c$ of the core-conduction ones.

Inserting Eqs.~\ref{eq7:reform1} and \ref{eq7:t2_def} into Eq.~\ref{eq:final}
yields
\begin{equation}\label{eq7:ddcs_final}
  \frac{\textrm{d}^2 \sigma}{\textrm{d} \Omega_2 \textrm{d}\omega_2}
  =\alpha^4 \left( \frac{\omega_2}{\omega_1} \right)
  \mathrm{Im} \sum_{\lambda_o}\frac{\left|\sum_{\lambda_c}
  \frac{t^{(2)}_{\lambda_o,\lambda_c}t^{(1)}_{\lambda_c}}
  {\omega_1 -E^{\lambda_c}+\mathrm{i}\eta}
  \right|^2}{(\omega_1-\omega_2)-E^{\lambda_o}+\mathrm{i}\eta}.
\end{equation}
Finally, we define the \textit{RIXS oscillator strength}
$t^{(3)}_{\lambda}(\omega_1)$ as
\begin{equation}\label{eq7:t3_def}
  t^{(3)}_{\lambda_o}(\omega_1)=\sum_{\lambda_c}
  \frac{t^{(2)}_{\lambda_o,\lambda_c}
  t^{(1)}_{\lambda_c}}{\omega_1-E^{\lambda_c}+\mathrm{i}\eta}.
\end{equation} 
Using the definition of the oscillator strength $t^{(1)}_{\lambda_c}$ in
Eq.~\ref{eq7:t1_def} and the excitation pathway $t^{(2)}_{\lambda_o,\lambda_c}$
in Eq.~\ref{eq7:t2_def}, allows for the compact expression of the RIXS cross
section as
\begin{equation}\label{eq7:ddcs_final2}
  \frac{\textrm{d}^2 \sigma}{\textrm{d} \Omega_2 \textrm{d}\omega_2}
  =\alpha^4 \left( \frac{\omega_2}{\omega_1} \right)
  \mathrm{Im}\sum_{\lambda_o}\frac{|t^{(3)}_{\lambda_o}(\omega_{1})|^2}
  {(\omega_1-\omega_2)-E^{\lambda_o}+\mathrm{i}\eta}
\end{equation}
which closely resembles the BSE expression for optical and x-ray absorption
spectra. The cross section depends explicitly on the energy loss
$\omega=\omega_1-\omega_2$, while its dependence on the excitation energy is
contained in the oscillator strength $t^{(3)}_{\lambda_o}(\omega_1)$. It has
poles in the energy loss at the optical excitation energies $E^{\lambda_o}$ of
the system, independent of the excitation energy, while the oscillator strength
of each of these excitations depends on it. The oscillator strength defined in
Eq.~\ref{eq7:t3_def}, gives further insight into the many-body processes that
occur in RIXS. The rate of the initial x-ray absorption event is given by
$t^{(1)}_{\lambda_c}$, combined with the energy conservation rule (the
denominator $\omega_1-E^{\lambda_c}+i\eta$ in Eq.~\ref{eq7:t3_def}). The
absorption leads to an intermediate core-excited state, characterized by the
excitation index $\lambda_c$. The final RIXS spectrum is then given by the rate
of the absorption combined with the pathway $t^{(2)}_{\lambda_o,\lambda_c}$ that
describes the many-body transition $|\lambda_c \rangle \rightarrow | \lambda_o
\rangle$. These pathways are far from obvious, as the mixing between
$t^{(1)}_{\lambda_c}$ and $t^{(2)}_{\lambda_o, \lambda_o}$ can develop in
destructive or constructive interference, attesting the many-body character of
such process.   

With Eq.~\ref{eq7:ddcs_final2}, we have derived a compact analytical expression
for the double-differential RIXS cross section with only two assumptions: First,
we presume in Eq.~\ref{eq:id_tda} that the intermediate and final states are
singlet excitations. While this assumption limits our approach to direct RIXS,
it is consistent with the Tamm-Dancoff approximation in the BSE formalism.
Second, we assume that the latter yield accurate core and valence excited
states. Therefore, these approximations are interconnected. For systems, such as
diamond and \ce{Ga2O3} studied here, where BSE yields accurate absorption
spectra, our approach yields accurate RIXS spectra as well. For highly
correlated states, BSE results may strongly depend on the starting point, \textit{i.e.}
the underlying one-particle states. It might happen, for instance, that
semilocal DFT leads to a poor representation of the system. In this case, even
the BSE result will be poor. This is neither a problem of BSE nor of the RIXS
formulation. Here, one-particle wave functions coming from hybrid functionals,
or self-consistent COHSEX or {\it GW} will be required, as shown for copper and
vanadium oxides.\cite{Bruneval2006,Gatti2015} Such calculations are, however,
not standard to date as they are numerically very involved. All in all, the RIXS
formulation provided here, is very general and fully {\it ab initio}, and it
applies to a large variety of systems.
%************************************************
\begin{figure}[t!]
  \centering
  \includegraphics[width=0.8\columnwidth]{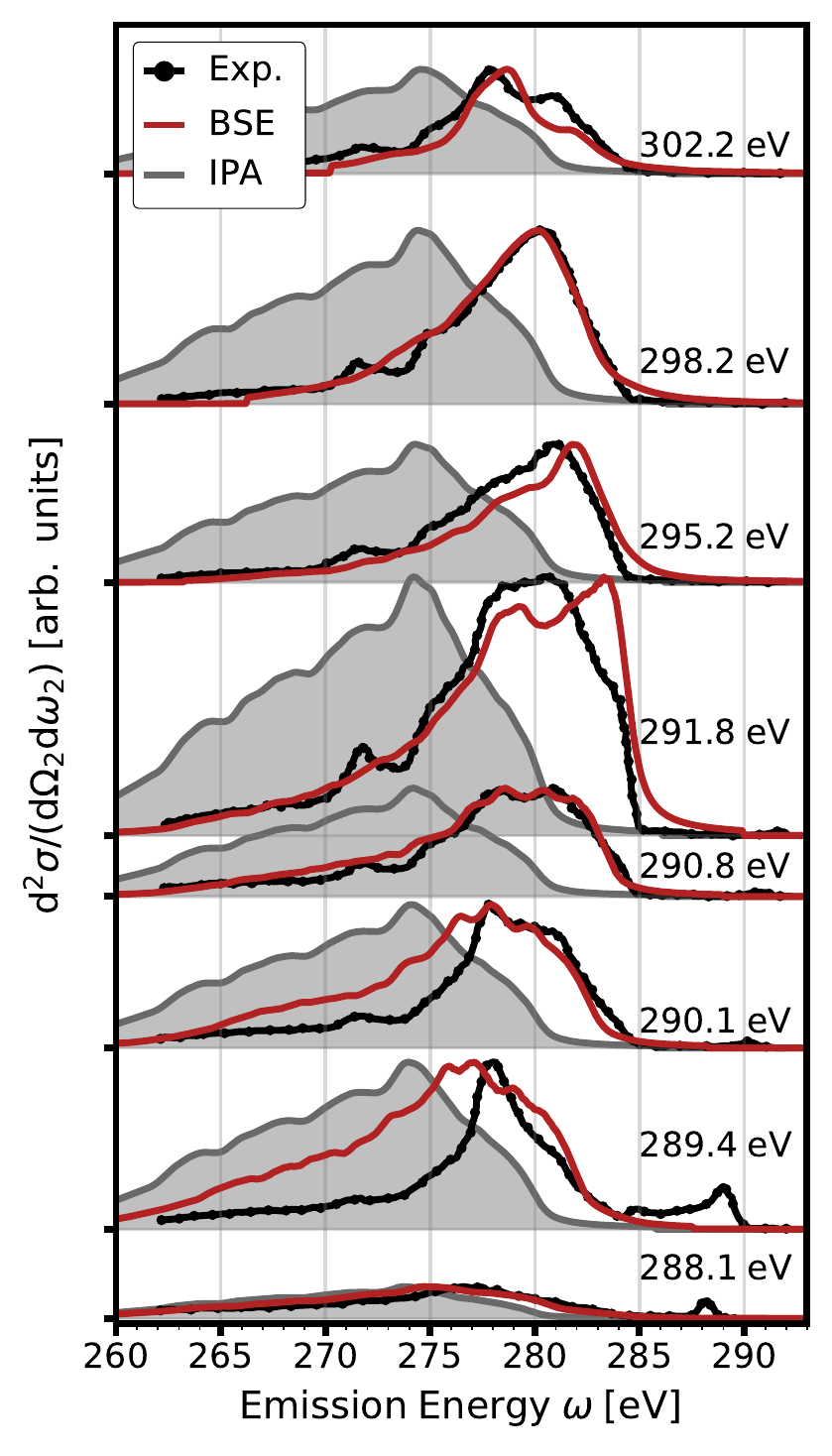}
  \caption{\label{fig:rixs-diamond2} Double differential RIXS cross section
(red) accounting for electron-hole interaction. The spectra calculated for
several excitation energies are offset for clarity. Experimental data from
Ref.~\cite{Sokolov-2003} are shown in black, the IPA results in gray. }
\end{figure}
%************************************************
%------------------------------------------------------------------------------
\section{Implementation}
%------------------------------------------------------------------------------
The implementation of Eq.~\ref{eq7:ddcs_final2} requires explicit access to the
BSE eigenvectors $X_{cv\mathbf{k} \lambda_o}$ both in terms of
valence-conduction transitions $X_{c \mu \mathbf{k}\lambda_c}$ as well as
core-conduction transitions, while cross terms of the form $X_{v \mu \mathbf{k}
\lambda}$ are not needed. Thus, the calculation of the RIXS cross section can be
separated into three independent calculations, one for the core-conduction, one
for the valence-conduction excitations, and finally a convolution step to obtain
the RIXS cross section. Overall, momentum matrix elements $\mathbf{P}_{c \mu
\mathbf{k}}$ between core and conduction states and $\mathbf{P}_{\mu v
\mathbf{k}}$ between valence and core states determine the excitation and
de-excitation process, respectively. The coherence between core and valence
excitations in the RIXS process is apparent in Eqs.~\ref{eq7:t2_def} from the
summation over $\mathbf{k}$.  (For this reason, core and valence excitations are
calculated on the same $\{ \mathbf{k} \}$-grid.) Coherence occurs, since the
absorption and emission processes conserve the crystal momentum
$\mathbf{k}$.\cite{Kotani2001}

For the specific implementation, we make use of the core-conduction and
valence-conduction BSE eigenvectors and corresponding matrix elements obtained
from the all-electron package
\texttt{exciting}.\cite{gulans2014,vorwerk2017,vorwerk2019} The output of the two BSE
calculations is evaluated by the \texttt{BRIXS} (\texttt{B}SE for \texttt{RIXS})
code.\cite{Vorwerk_BRIXS_2021,vorwerk2020} In this step, the oscillator strength
$t^{(1)}$ of the core excitation and the excitation pathways $t^{(2)}$ are
determined. From these intermediate quantities, the RIXS oscillator strength
$t^{(3)}(\omega_1)$ is generated for a list of excitation energies $\omega_1$
defined by the user. Execution of \texttt{BRIXS} requires only a minimal number
of input parameters: Besides $\{ \omega_1\}$, the number of core-conduction
($N_{\lambda_c}$) and valence-conduction excitations ($N_{\lambda_o}$), the
lifetime broadening ($\eta$), and the polarization vector of the x-ray beam
($\mathbf{e}_1$) are required. A detailed description of the implementation is
provided in Ref. \cite{vorwerk2021}

\section{Results}
%------------------------------------------------------------------------------
\subsection{Electron-hole Correlation in the RIXS of Diamond}
%------------------------------------------------------------------------------
To demonstrate the importance of electron-hole correlation in RIXS, we study the
carbon K edge in diamond. Both optical
absorption\cite{phillip1964,benedict1998,Hahn2005,hanke1974,rocca2012} and core
excitations \cite{Sokolov-2003,soininen2001,taillefumier2002} have been
investigated intensively before. RIXS measurements \cite{Ma1992,Sokolov-2003}
and calculations \cite{shirley2000,johnvinson2012} are also available for this
edge. As such, this material acts as a good example to demonstrate our approach
and benchmark the resulting spectra.

In a first step, the optical and core spectra are calculated from the solution
of the BSE, as shown in Fig.~\ref{fig:rixs-diamond1}, both in excellent
agreement with experiment. In addition, we show the results obtained within the
independent-particle approximation (IPA). They are blue-shifted relative to
experiment, as electron-hole attraction is not included. Also the spectral shape
of the core spectra disagrees with the experimental one, as it misses the peak
at the absorption onset and shows too much spectral weight at high energies.

From the BSE spectra, we determine the RIXS specrtra shown in
Fig.~\ref{fig:rixs-diamond1}. Following Eqs.~\ref{eq7:ddcs_final} and
\ref{eq7:t3_def}, it comes natural to display the RIXS double-differential cross
section $\mathrm{d}^2 \sigma/\mathrm{d} \Omega_2 \mathrm{d} \omega_2$ as a
function of the excitation energy $\omega_1$ and the energy loss
$\omega=\omega_1-\omega_2$. For excitation energies below the core absorption
edge, \textit{i.e.} at approximately 290 eV (see Fig.~\ref{fig:rixs-diamond2}, right), the
intensity is negligible, since the excitation energy is not in resonance with
any carbon $1s$ excitation. Once the excitation energy reaches resonance with
the absorption edge, the RIXS cross section increases considerably. The emission
occurs over a wide range of the energy loss up to around 20 eV, but is strongest
at low emission at the onset of optical absorption (compare
Fig.~\ref{fig:rixs-diamond1}, top). With increasing excitation energy, the
emission reduces due to the reduced rate of absorption beyond the onset.
Furthermore, the emission at low energy loss vanishes as the excitation energy
increases. At a value of about 295\,eV (300\,eV), no emission with an energy
loss below approximately 12\,eV (20\,eV) is observed. Due to this linear
dispersion of the energy loss with the excitation energy, the emission energies
stay more or less constant, as can be seen in Fig.~\ref{fig:rixs-diamond2},
where the RIXS cross section is shown as a function of the emission energy for
selected excitation energies.
%************************************************
\begin{figure}[t!]
  \centering
  \includegraphics[width=0.9\columnwidth]{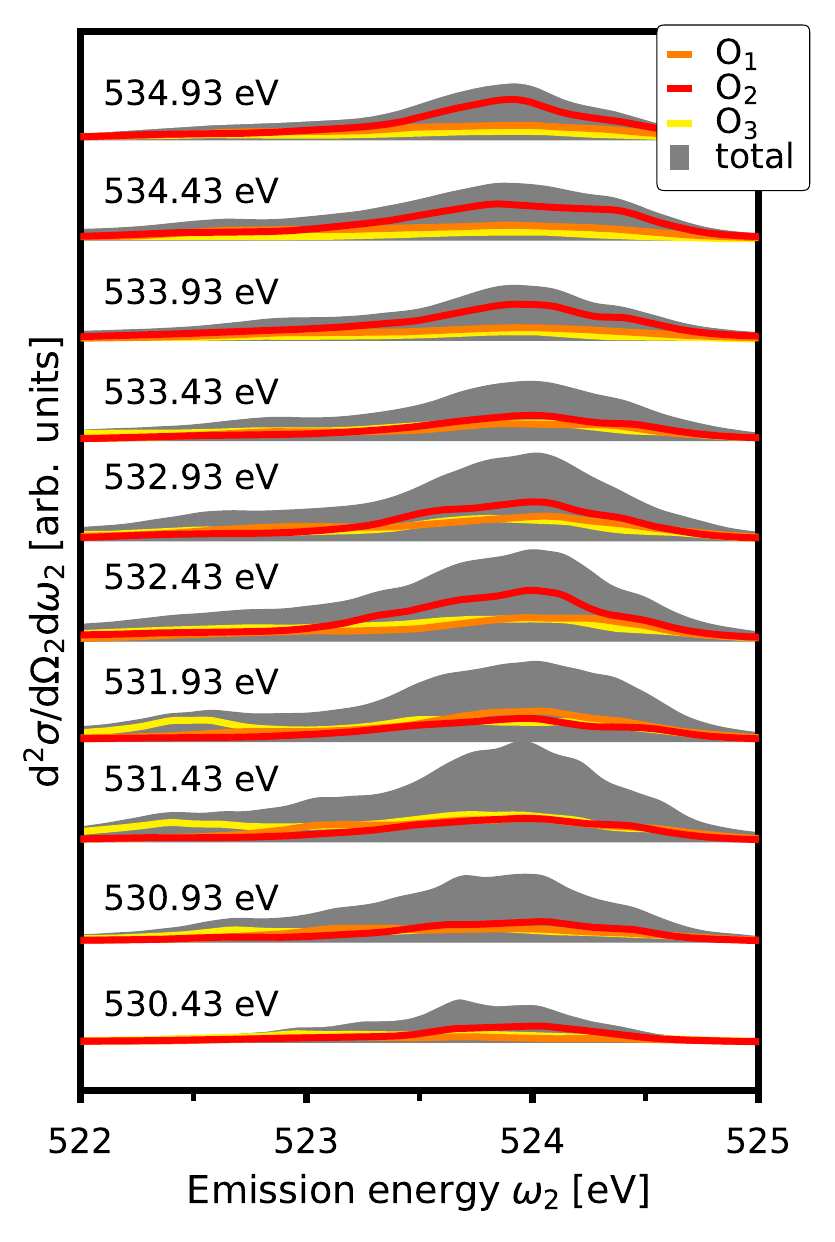}
  \caption{\label{fig:rixs-ga2o3}Total O K edge RIXS in $\beta$-\ce{Ga2O3} as a
function of emission energy, $\omega_2$, for selected excitation energies,
together with the contributions from $\mathrm{O}_1$, $\mathrm{O}_2$, and
$\mathrm{O}_3$. The spectra are offset for clarity.} 
\end{figure}
%************************************************

In Fig.~\ref{fig:rixs-diamond2}, we also show the RIXS cross section obtained
within the IPA, following Eq.~\ref{eq4:rixs7-1}. Overall, the IPA spectra
exhibit lower emission energies due to the overestimation of the excitation
energies in the optical absorption (see Fig.~\ref{fig:rixs-diamond1}). 
Especially for lower excitation energies, there are significant deviations. The
intensity at higher emission energies is underestimated, and the spectra are too
broad. Both features are due to the neglect of electron-hole interaction which
increases the intensity at low energies, \textit{i.e.} leads to a sharper absorption
onsets.  This effect is especially pronounced for the core excitations. At
higher excitation energies, the discrepancy between IPA and BSE results
decreases. While IPA reproduces the spectral shape correctly, the spectrum is
still blue-shifted. Our BSE results are in excellent agreement with experiment
for all excitations energies, both in relative position and
spectral shape.

%------------------------------------------------------------------------------
\subsection{Atomic Coherence in the RIXS of \ce{Ga2O3}}
\begin{figure}[t!]
  \centering
  \includegraphics[width=0.9\columnwidth]{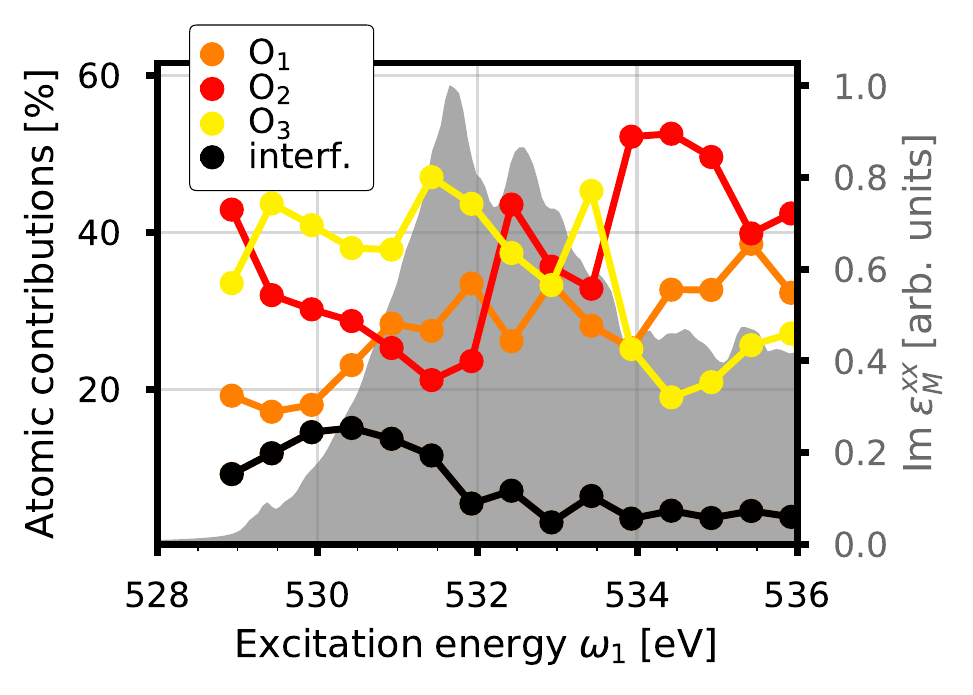}
  \caption{\label{fig11:ga2o3-ok-rixs-b3}Interference contribution to the O K
edge RIXS of $\beta-\ce{Ga2O3}$ as a function of excitation energy compared to
the atomic contributions and the normalized total spectrum (gray). } 
\end{figure}

%------------------------------------------------------------------------------
Now, we like to showcase the importance of atomic coherence on RIXS spectra.
Considering the coherent sum over atomic excitations in
Eq.~\ref{eq7:ddcs_final}, interference terms appear when the crystal contains
inequivalent atomic positions, \textit{i.e.}
\begin{equation}
  \frac{\mathrm{d}^2 \sigma}{\mathrm{d}\Omega_2 \mathrm{d}\omega_2}=
  \sum_{a}^{N_{\mathrm{atoms}}}M_{a}^2 \frac{\mathrm{d}^2 \sigma_{a}}
  {\mathrm{d}\Omega_2 \mathrm{d}\omega_2}+
  \frac{\mathrm{d}^2 \sigma_{\mathrm{interf}}}
  {\mathrm{d}\Omega_2 \mathrm{d}\omega_2},
\end{equation}
where $\mathrm{d}^2 \sigma_a / \mathrm{d}\Omega_2 \mathrm{d}\omega_2$ is the
RIXS cross section for the inequivalent atom $a$ with multiplicity $M_a$ and
$\mathrm{d}^2 \sigma_{\mathrm{interf}}/\mathrm{d}\Omega_2 \mathrm{d} \omega_2$
is the interference term. Mathematically, the interference term originates from
the square modulus of $t^{(3)}_{\lambda}(\omega_1)$ in Eq.~\ref{eq7:ddcs_final}
that contains the sum of the core excitations on all atoms in the unit cell. The
O K edge in the monoclinic $\beta$ phase of \ce{Ga2O3} serves as an example. The
unit cell contains three inequivalent oxygen sites denoted $\mathrm{O}_1$,
$\mathrm{O}_2$, and $\mathrm{O}_3$, all of them with multiplicity 1. The atomic
positions are provided in the SI. Figure \ref{fig:rixs-ga2o3}  shows that the
RIXS spectrum shows pronounced fluorescence behavior,\cite{pfaff2018,huang2016}
\textit{i.e.} significant features occur at basically constant emission energies. The
contributions of all oxygen atoms are nearly identical for excitation energies
below 532 eV. Above that, the contribution of $\mathrm{O}_2$ dominates the
spectrum. For a quantitative analysis, we define relative atomic contributions
by integrating the atomic RIXS spectra over the emission energies $\omega_2$ and
normalizing with respect to the total spectrum. The result, shown in
Fig.~\ref{fig11:ga2o3-ok-rixs-b3}, demonstrates that the relative atomic
contributions vary between 20 and 50\%, depending on the excitation energy, and
that none of the contributions can be neglected even if one of them -- here
$\mathrm{O}_2$ -- dominates. The interference term contributes up to 15\% of the
total RIXS spectrum, yet decreasing quickly as the excitation energy increases
beyond the absorption onset. At higher excitation energies, it only contributes
5-7\%. The importance of interference at the absorption onset indicates that it
originates from electronic correlation. 

%************************************************
%************************************************

%------------------------------------------------------------------------------
\section{Conclusions}
%------------------------------------------------------------------------------
We have derived a compact analytical expression for the RIXS cross section
within many-body perturbation theory.  This expression retains the intuitive
interpretation of the RIXS process as a coherent absorption-emission process,
while including the effects of electron-hole correlation, which are paramount
for an accurate description of excitations in semiconductors and insulators. Our
implementation in an all-electron BSE framework, \textit{i.e.} the
\texttt{exciting} package,
making use of BSE eigenstates for core and valence excitations, introduces only
small computational overhead compared to valence and core BSE calculations.  For
the example of the carbon K edge RIXS in diamond, we demonstrate that our
approach yields spectra in excellent agreement with available experimental
spectra. We furthermore show that electron-hole correlations not only shift
spectral features in energy but also affect their shape, especially in
excitations at the absorption edge. The influence of atomic coherence is
exemplified with the oxygen K edge in $\beta$-\ce{Ga2O3}, where it turns out
non-negligible, especially close to the absorption onset.

%------------------------------------------------------------------------------
\section*{Author Contributions}
%------------------------------------------------------------------------------
All authors have devised the project. FS and CV derived the RIXS formalism, CV
implemented it in the all-electron package \texttt{exciting} and carried out the
calculations. All authors contributed to the writing of the manuscript.

%------------------------------------------------------------------------------
\section*{Conflicts of interest}
%------------------------------------------------------------------------------
There are no conflicts to declare.

%------------------------------------------------------------------------------
\section*{Acknowledgements}
%------------------------------------------------------------------------------
This work was supported by the project GraFOx, a Leibniz ScienceCampus,
partially funded by the Leibniz Association. FS thanks the French Agence
Nationale de la Recherche (ANR) for financial support (grant No.
ANR-19-CE30-0011).
%------------------------------------------------------------------------------
\section*{Appendix}
%------------------------------------------------------------------------------
%------------------------------------------------------------------------------
\subsection*{Derivation of Matrix Elements}
%------------------------------------------------------------------------------
In the following, we provide the derivation of Eq.~\ref{eq:rixs7}. We initially
employ Wick's theorem~\cite{Fetter_Book} to evaluate the expectation value of
the product of three pairs of annihilation and creation operators, \textit{i.e.}
\begin{equation}\label{eq:wick}
  \begin{aligned}
    \langle c v \mathbf{k} | \hat{c}_{m \mathbf{k}''}^{\dagger}
    \hat{c}_{n \mathbf{k}''} |c' \mu \mathbf{k}' \rangle &=
    \langle 0 |\left[ \hat{c}^\dagger_{v \mathbf{k}} \hat{c}_{c \mathbf{k}}
    \right] 
    \left[ \hat{c}^\dagger_{m \mathbf{k}''} \hat{c}_{n \mathbf{k}''} \right]
    \left[ \hat{c}^\dagger_{c' \mathbf{k}'} \hat{c}_{\mu \mathbf{k}'}
    \right]| 0 \rangle\\
    &= \delta_{v \mu} \delta_{c'n}\delta_{cm}\delta_{\mathbf{k}\mathbf{k}'}
    \delta_{\mathbf{k}\mathbf{k}''}\\
    &+\delta_{cc'} \delta_{mn}\delta_{v \mu}\delta_{\mathbf{k}\mathbf{k}'}
    \delta_{\mathbf{k}\mathbf{k}''}\\
    &-\delta_{cc'} \delta_{\mu m}\delta_{vn}\delta_{\mathbf{k}\mathbf{k}'}
    \delta_{\mathbf{k}\mathbf{k}''},
  \end{aligned}
\end{equation}
Each of the three contributions correspond to different processes in the DDCS.
To analyze them, we insert Eq.~\ref{eq:wick} into Eq.~(7). Ignoring cross-terms,
this yields the following expressions:
\begin{equation}\label{eq:rixsIP}
  \begin{aligned}
    &\frac{\textrm{d}^2 \sigma}{\textrm{d}\Omega_2\textrm{d}\omega_2} \bigg|_{IP}
  \!\!\!\!\!\!=\\
    &\alpha^4 \left(\frac{\omega_2}{\omega_1}\right)\!\!
  \begin{cases} 
    \begin{array}{c} \sum_{c' \mu \mathbf{k}}\left|\sum_{c}
    \frac{\mathbf{e}_2^* \cdot \mathbf{P}_{c' c \mathbf{k}}
    \mathbf{P}_{c \mu \mathbf{k}}\cdot \mathbf{e}_1}
    {\omega_1-(\epsilon_{c \mathbf{k}}-\epsilon_{\mu \mathbf{k}})+\mathrm{i}\eta}
    \right|^2 \times \\
    \times \delta(\omega-(\epsilon_{c' \mathbf{k}}-\epsilon_{\mu \mathbf{k}}))
    \end{array}
    & \textrm{(a)}\\
    \begin{array}{c}
      +\sum_{c \mu \mathbf{k}}\left|\sum_{m} 
      \frac{\mathbf{e}^*_2 \cdot \mathbf{P}_{mm \mathbf{k}}
      \mathbf{P}_{c \mu \mathbf{k}}\cdot \mathbf{e}_1}
      {\omega_1-(\epsilon_{c \mathbf{k}}-\epsilon_{\mu \mathbf{k}})+\mathrm{i}\eta}
      \right|^2 \times \\ 
      \times \delta(\omega-(\epsilon_{c \mathbf{k}}-\epsilon_{\mu \mathbf{k}}))
    \end{array}
    & \textrm{(b)}\\
    \begin{array}{c}
      +\sum_{cv \mathbf{k}}\left|\sum_{\mu} 
      \frac{\mathbf{e}^*_2 \cdot \mathbf{P}_{\mu v \mathbf{k}}
      \mathbf{P}_{c \mu \mathbf{k}}\cdot \mathbf{e}_1}
      {\omega_1-(\epsilon_{c \mathbf{k}}-\epsilon_{\mu \mathbf{k}})+\mathrm{i}\eta}
      \right|^2 \times \\
      \times \delta(\omega-(\epsilon_{c \mathbf{k}}-\epsilon_{v \mathbf{k}}))
    \end{array}
    & \textrm{(c)}
  \end{cases}
  \end{aligned}
\end{equation}
An intuitive interpretation of the three terms is obtained by considering their
poles: Term (a) has poles at the excitation energies $\omega_1 = \epsilon_{c
\mathbf{k}}-\epsilon_{\mu \mathbf{k}}$, where a core state $\mu \mathbf{k}$ is
excited into a conduction state $c \mathbf{k}$. Poles in the emission energy
occur at $\omega_2 = \epsilon_{c' \mathbf{k}}- \epsilon_{c \mathbf{k}}$ where
the excited electron transitions to another conduction state $c' \mathbf{k}$.
Thus, the final state of this scattering process contains a core hole at $\mu
\mathbf{k}$ and an excited electron in $c' \mathbf{k}$. As such, this scattering
process does not correspond to the RIXS process.

Term (b) does not describe a resonant scattering process, as the emission energy
$\omega_2$ vanishes (corresponding to transitions $m \mathbf{k} \rightarrow m
\mathbf{k}$).  However, for each state, the momentum matrix element
$\mathbf{P}_{mm \mathbf{k}}=0$. Thus, term (b) does not contribute to the DDCS.

Finally, term (c) corresponds to the RIXS process: Poles occur at excitation
energies $\omega_1=\epsilon_{c \mathbf{k}}-\epsilon_{\mu \mathbf{k}}$ and at
emission energies $\omega_2 = \epsilon_{v \mathbf{k}} - \epsilon_{\mu
\mathbf{k}}$. Thus, the scattering consists of an excitation $\mu \mathbf{k}
\rightarrow c \mathbf{k}$ and a subsequent de-excitation $v \mathbf{k}
\rightarrow \mu \mathbf{k}$. The final state contains a valence hole $v
\mathbf{k}$ and an
excited electron $c \mathbf{k}$.

Overall, as long we only consider RIXS, we can neglect terms (a) and (b), and we
arrive at Eq.~\ref{eq:final}. The analysis also justifies why the cross terms
between the terms (a), (b), and (c) can be neglected. The different terms have
poles in vastly different energy regions, thus making cross terms small.

%------------------------------------------------------------------------------
\subsection*{Numerical Parameters}
%------------------------------------------------------------------------------
%------------------------------------------------------------------------------
\subsubsection*{Carbon K Edge in Diamond}
%------------------------------------------------------------------------------
The electronic structure is determined from DFT-PBE calculations within the
\texttt{exciting} code. All calculations are performed for the experimental
lattice parameter of $6.746 \;a_0$. The reciprocal space is sampled with a $ 9
\times 9 \times 9$ $\mathbf{k}$-grid, and we include basis functions up to a
cut-off of $R_{MT}^{max}\cdot |\mathbf{G}+\mathbf{q}|_{max}=8$. To correct the
electronic structure, we apply a scissors operator $\Delta \omega=1.9$ eV, such
to reproduce the measured indirect band gap of 5.48 eV.\cite{clark1964} Another
scissors operator of 22 eV is employed to correct the position of the $1s$
level.

Optical BSE spectra are calculated on a $13 \times 13 \times 13$
$\mathbf{k}$-grid, and local-field effects are included up to a cut-off
$|\mathbf{G}+\mathbf{q}|_{\mathrm{max}} = 3.5 \;a_0^{-1}$. The calculations
include all 4 valence bands and 10 conduction bands. 100 conduction bands are
included in the random-phase approximation (RPA) calculation to obtain the
screened Coulomb potential. The carbon K edge BSE is performed on the same
$\mathbf{k}$-grid. The cut-off for local-field is chosen
$|\mathbf{G}+\mathbf{q}|_{\mathrm{max}}=5.5 \; a_0^{-1}$ such to provide a
precise description of the more localized excitations. 40 unoccupied are
included in the BSE calculation. For the \texttt{BRIXS} calculation, the 8.000
lowest-energy valence and 20.000 carbon $1s$ excitations are taken into account.

Input and relevant output files of the electronic-structure and BSE calculations
can be downloaded from the NOMAD Repository\cite{Draxl_2018,Draxl2019} under the
DOI provided in Ref. \cite{NOMADdiamond}.

%------------------------------------------------------------------------------
\subsubsection*{O K Edge in $\beta$-\ce{Ga2O3}}
%------------------------------------------------------------------------------
Calculations for $\beta$-\ce{Ga2O3} are performed using the experimental lattice
parameters $a=12.233 \; \si{\angstrom}$, $b=3.038\; \si{\angstrom}$, $c=5.807 \;
\si{\angstrom}$, and $\beta=103.82^{\circ}$.\cite{lipinska-kalita2008} The
electronic structure is determined on a $8 \times 8 \times 8$ $\mathbf{k}$-grid
using basis functions up to a cut-off of $R _{MT}^{max}\cdot
|\mathbf{G}+\mathbf{q}|_{max}=8$. Our calculations with the PBEsol
functional~\cite{perdew2008} yield a Kohn-Sham gap of 2.89 eV. To match the
experimental fundamental gap of of 5.72 eV,\cite{segura2017} we apply a scissors
operator of $\Delta \omega =2.6$ eV. A scissors shift of $\Delta \omega_2 =
24.93$ eV is applied to correct the position of the oxygen $1s$ state such to
align the calculated O K edge XAS with the experimental one.\cite{swallow2020}

The optical BSE spectra are obtained from calculations on a $10 \times 10 \times
10$ $\mathbf{k}$-grid with a cut-off $|\mathbf{G}+\mathbf{q}|_{max} = 1.1 \;
a_0^{-1}$ for local-field effects. The 10 highest valence bands and 10 lowest
conduction bands form the transition space. 30 conduction bands are used in the
RPA calculation of the screened Coulomb potential.

The BSE calculation of the oxygen K edge are performed on the same
$\mathbf{k}$-grid and using the identical screened Coulomb potential as for the
optical spectra. 20 conduction bands are used to form the transition space.

Input and relevant output files of the electronic-structure and BSE calculations
can be downloaded from the NOMAD Repository\cite{Draxl_2018,Draxl2019} under the
DOI provided in Ref.~\cite{NOMADbeta}

%%%REFERENCES%%%
\bibliography{bibliography} %You need to replace "rsc" on this line with the name of your .bib file

\end{document}